\documentclass[fleqn,10pt]{wlscirep}

\usepackage{pdfpages} 
\usepackage{pgffor} 

\usepackage{gensymb}
\usepackage{graphicx}
\usepackage[caption=false]{subfig}
\usepackage{tikz}
\usepackage{tabularx}
\usepackage{cite}
\usepackage{url}
\usepackage{dirtree}

\usepackage{verbatim}
\usepackage{longtable}

\usepackage{booktabs}
\usepackage{amsmath}
\usepackage[version=3]{mhchem}
\usepackage{siunitx}

\usetikzlibrary{shapes.geometric, arrows}
\usetikzlibrary{positioning}
\usetikzlibrary{calc}
\usepackage{standalone}
\usepackage{multirow}
\usepackage{booktabs}

\usepackage{rotating}
\usepackage{arydshln}
\usepackage{hhline}
\usepackage{makecell}
\usepackage{bm}

\usepackage{pgfplots}
\usepackage{amssymb}

\usetikzlibrary{arrows,positioning}
\usetikzlibrary{plotmarks}
\usetikzlibrary{calc}
\usetikzlibrary{shapes.geometric}

\pgfplotsset{compat=1.8}
\usepgfplotslibrary{statistics}
\usepgfplotslibrary{groupplots}
\usetikzlibrary{shapes,arrows,positioning, fit}

\tikzstyle{block} = [rectangle, rounded corners, minimum height=1cm, text centered, draw=black]

\newcommand{\benchmark}{MuS2}
\newcommand{\dice}{D_C}

\newcommand{\srout}{\mathbb{I}_{\rm SR}}
\newcommand{\hrref}{\mathbb{I}_{\rm HR}}
\newcommand{\lrinput}{\mathbb{I}_{\rm in}}

\newcommand{\bicout}{\mathbb{I}_{\rm bic}}

\usepackage{lineno}

\title{MuS2: A Real-World Benchmark for Sentinel-2 Multi-Image Super-Resolution}

 \author[1,2]{Pawel Kowaleczko}
 \author[3]{Tomasz Tarasiewicz}
 \author[1]{Maciej Ziaja}
 \author[1,3]{Daniel Kostrzewa}
 \author[1,3]{Jakub~Nalepa}
 \author[2]{Przemyslaw~Rokita}
 \author[1,3,*]{Michal~Kawulok}
\affil[1]{KP Labs, Gliwice, Poland}
\affil[2]{Warsaw University of Technology, Warsaw, Poland}
\affil[3]{Silesian University of Technology, Faculty of Automatic Control, Electronics and Computer Science, Gliwice, Poland}

 \affil[*]{corresponding author(s): Michal Kawulok (michal.kawulok@polsl.pl)}

\begin{abstract}
Insufficient image spatial resolution is a serious limitation in many practical scenarios, especially when acquiring images at a finer scale is infeasible or brings higher costs. This is inherent to remote sensing, including Sentinel-2 satellite images that are available free of charge at a high revisit frequency, but whose spatial resolution is limited to 10\,m ground sampling distance. The resolution can be increased with super-resolution algorithms, in particular when performed from multiple images captured at subsequent revisits of a satellite, taking advantage of information fusion that leads to enhanced reconstruction accuracy.
One of the obstacles in multi-image super-resolution consists in the scarcity of real-world benchmarks---commonly, simulated data are exploited which do not fully reflect the operating conditions. In this paper, we introduce a new MuS2 benchmark for super-resolving multiple Sentinel-2 images, with WorldView-2 imagery used as the high-resolution reference. Within MuS2, we publish the first end-to-end evaluation procedure for this problem which we expect to help the researchers in advancing the state of the art in multi-image super-resolution.
\end{abstract}

\begin{document}

\flushbottom
\maketitle
\thispagestyle{empty}


\section*{Background \& Summary}

Super-resolution (SR) is aimed at reconstructing a high-resolution (HR) image from a single image or multiple low-resolution (LR) observations presenting the same scene. Multi-image SR (MISR) fuses multiple LR images, each of which contains a different portion of HR information. This allows for achieving higher reconstruction accuracy than relying on single-image SR (SISR)~\cite{Yue2016}, but MISR is highly sensitive to the variability of the input images and their proper co-registration~\cite{Molini2020}. This poses a challenge when preparing the data for training and validation. 
Recent advances in satellite image SR include SISR~\cite{Galar2020} and MISR~\cite{Razzak2021} techniques for enhancing Sentinel-2 (S-2) multispectral images (MSIs), composed of 13 bands, whose resolution ranges from 60\,m ground sampling distance (GSD) to 10\,m GSD~\cite{Drusch2012}.



Commonly, SR techniques are evaluated relying on an artificial scenario---a certain image is treated as an HR reference which is subsequently degraded to obtain the simulated LR images. The similarity of the super-resolved outcome to the reference is then used to evaluate the SR performance. Unfortunately, such procedure does not reflect the real-world operating conditions~\cite{ChenHe2022}, and methods that perform well for the simulated data are not necessarily effective for original (i.e., not downsampled) images.
It is therefore crucial to properly validate the emerging techniques using real LR images coupled with a real HR reference---in an excellent survey on real-world SISR~\cite{ChenHe2022}, Chen et al. identified the deficiency of realistic datasets as one of the most important challenges in this field. Recently, several real-world SISR datasets have been elaborated~\cite{Joze2020ImagePairs,Wei2020DRealSR,Bhat2021NTIRE}, 
however preparing such datasets for MISR is much more costly and troublesome.

In 2019, European Space Agency organized an SR challenge~\cite{Martens2019} based on real-world scenes acquired by the Proba-V satellite, each of which contains an HR image (100\,m GSD) coupled with at least nine LR images (300\,m GSD). The dataset allowed for developing first MISR techniques underpinned with convolutional neural networks (CNNs), applied either to enhance the LR images before their multi-temporal fusion~\cite{Kawulok2020GRSL} or employed to learn the reconstruction process in an end-to-end manner~\cite{Molini2020,Valsesia2022,AnZhang2022}.
Very recently, the WorldStrat dataset was published which matches multiple S-2 images with SPOT spectral bands of 6\,m GSD~\cite{Cornebise2022} (the magnification factor equals $1.67\times$). However, the problem of comparing the reconstruction outcome with the reference was not discussed there and it is not clear whether and how WorldStrat can be used for benchmarking SR algorithms. We demonstrate that this is not a straightforward task, especially when LR and HR images are captured by different satellites. In~\cite{Beaulieu2018}, S-2 SISR was assessed by comparing the outcome against WorldView-3 images. The authors observed that the peak signal-to-noise ratio (PSNR) and structural similarity index (SSIM) do not correlate well with the quality of the reconstructed images---the highest scores were obtained for blurred images in which the details were not reconstructed accurately. Contrary to that, the learned perceptual image patch similarity (LPIPS)~\cite{ZhangIsola2018} was reported to be suitable for evaluating SISR for remote sensing~\cite{WangBayram2022,DongZhang2022}, but it was not applied to MISR.

\begin{figure}[!t]
\centering
\includegraphics[width=\columnwidth]{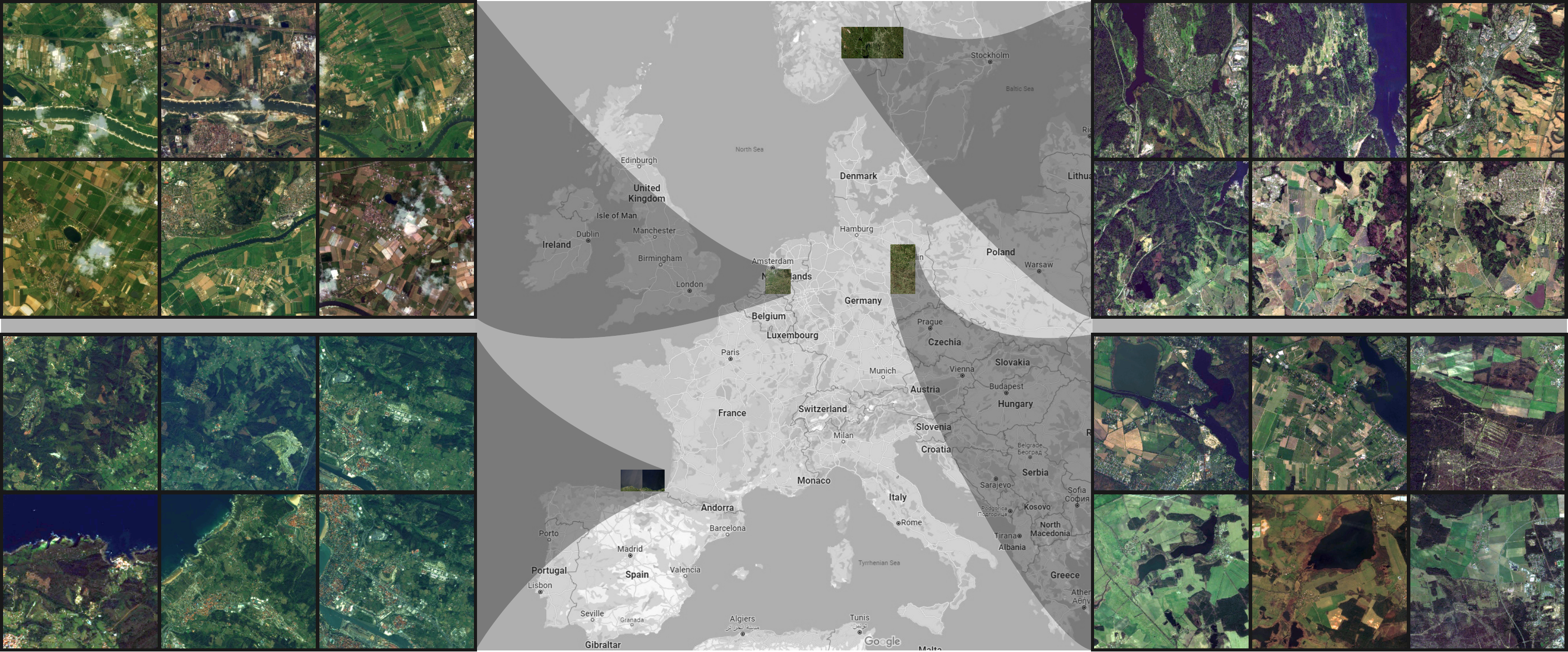}
\caption{Regions selected for our \benchmark\, benchmark (map source: Google).}
\label{fig:map}
\end{figure}

In this paper, we address the problem of evaluating MISR for S-2 images by introducing a new \underline{Mu}lti-image \underline{S}entinel-\underline{2} SR (\benchmark) benchmark. It is composed of a new dataset with WorldView-2 (WV-2) images used as an HR reference and the end-to-end validation procedure. Our contribution can be summarized in the following points.
\begin{itemize}
    \item We publish a new \benchmark\, benchmark dataset with 91 diverse scenes covering around 2500\,km$^2$ (Fig.~\ref{fig:map}), each composed of at least 14 S-2 MSIs coupled with a single WV-2 MSI. The benchmark is intended to serve as a test set for evaluating future advancements in MISR. 
    \item We elaborated a protocol (included into the source code we publish) for assessing the SR outcome for all of S-2 10\,m bands based on the corresponding WV-2 bands, and we report the results of $3\times$ magnification obtained using the well-established MISR techniques.
    \item We demonstrate that the LPIPS metric~\cite{ZhangIsola2018} is suitable for evaluating the MISR outcome, even if the reference HR images are acquired using a different satellite. To verify that, we have performed a mean opinion score (MOS) survey, whose results are thoroughly discussed.
    \item We introduce relevance masks which indicate the image regions that can be considered for evaluation relying on pixel-wise image similarity metrics.
\end{itemize}

\section*{Methods}

\subsection*{Data Source} \label{sec:datasource}
Our benchmark dataset is composed of images acquired by S-2 and WV-2 satellites. The S-2 images were downloaded from the Copernicus Open Access Hub, and we accessed the WV-2 images published within the European Cities dataset\footnote{The European Cities dataset is freely available at \url{https://earth.esa.int/web/guest/-/worldview-2-european-cities-dataset}.}. The S-2 data are organized into tiles, each of which contains a $100\times100$~km MSI, 
and aggregated into products that are distributed at different processing levels. Here, we exploit Level-2A which includes the bottom of atmosphere reflectance correction. Each tile is composed of 13 spectral bands
, however the B10 (cirrus) band is excluded from Level-2A, because it does not represent any surface information. The blue, green, red, and near infrared (NIR) bands (B02, B03, B04, and B08, respectively) are of 10\,m GSD, the vegetation red edge bands (B05, B06, and B07), narrow NIR (B08a), and the short-wave infrared (SWIR) bands (B11 and B12) are of 20\,m GSD, while the remaining coastal aerosol (B01), water vapour (B09), and cirrus clouds estimation (B10) bands are of 60\,m GSD.

Each WV-2 tile contains a panchromatic image of 0.4\,m GSD and 8 spectral bands at 1.6\,m GSD. These are C---coastal (400--450\,nm), B---blue (450--510\,nm), G---green (510--580\,nm), Y---yellow (585--625\,nm), R---red (630--690\,nm), RE---red edge (705--745\,nm), NIR1 (770--895\,nm), and NIR2 (860--1040\,nm).
Based on the coverage between the spectral ranges of S-2 and WV-2 bands (measured with the Dice coefficient---$\dice$), we have selected four pairs: B02 with B ($\dice=0.8$), B03 with G ($\dice=0.68$), B04 with R ($\dice=0.68$), and B08 with NIR1 ($\dice=0.92$). For the remaining band pairs, the spectral coverage did not exceed $\dice=0.5$, so we do not consider them in our study.

\subsection*{Scene Selection and Data Alignment}
\begin{figure}
    \centering
    \includegraphics[width=0.75\columnwidth]{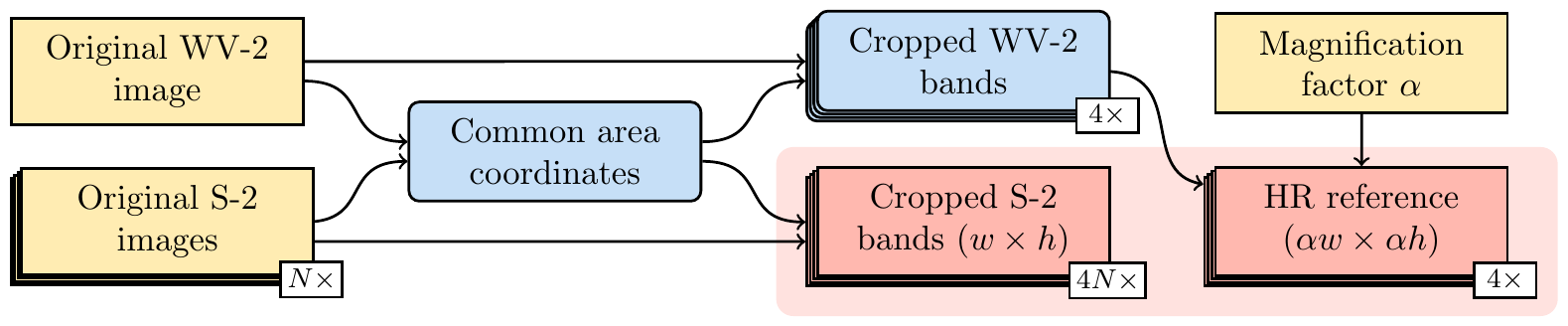}
    \caption{The data preparation pipeline: the yellow boxes indicate input data, the blue ones---temporary data, and the red ones---the output. For each WV-2 image, $N$ S-2 images with overlapping area are processed to identify the coordinates of the largest common region. It is used to crop the four 10\,m  bands from S-2 images and corresponding WV-2 bands. The latter are downsampled, so that they are $\alpha \times$ larger than the cropped S-2 images. }
    \label{fig:flowchartA}
\end{figure}

For each WV-2 tile, we collected 14 or 15 S-2 images that entirely contain that tile or overlap with it. The images, both within the S-2 stacks, as well as with the WV-2 tiles, were co-registered at whole-pixel precision.
We assemble the dataset using areas defined by seven different military grid reference system (MGRS) tiles. These areas represent diverse environments throughout Europe, such as mountains in Norway, plains in Germany and Belgium and a coastal region in Spain. The diversity of the chosen locations is illustrated in Fig.~\ref{fig:map}. Our data preparation procedure (see Fig.~\ref{fig:flowchartA}) operates on original S-2 and WV-2 images. The common area is determined based on the geographic coordinates retrieved from the metadata, and for each band, $N$ LR images ($\lrinput$) are cropped and coupled with the cropped WV-2 image. The latter is subsequently downsampled to create the HR reference ($\hrref$) whose each dimension is $\alpha \times$ larger than those of $\lrinput$. 
In the published dataset, we use $\alpha=3$, as enlarging the images $3\times$ remains sufficiently challenging for the state-of-the-art MISR techniques~\cite{AnZhang2022,deudon2020highresnet,Salvetti2020,Valsesia2022}. However, the elaborated software tools that we publish along with the benchmark allow for generating the reference HR data up to $\alpha=6.25$ for WV-2 images of original size.

\begin{figure}
    \centering
    \includegraphics[width=\columnwidth]{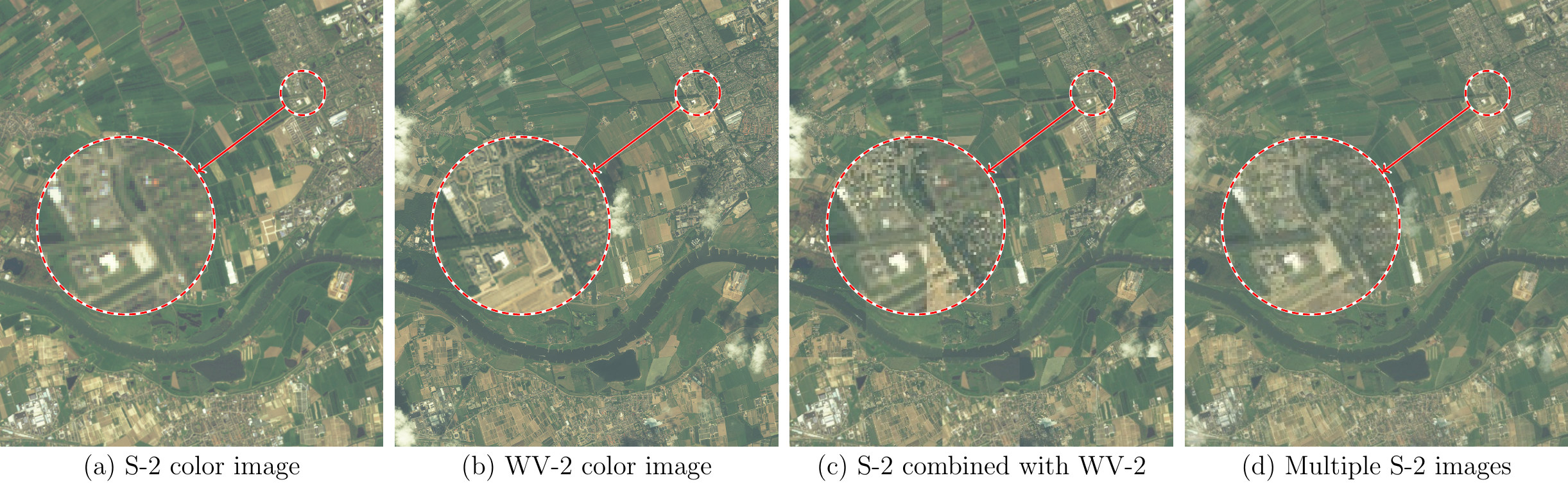}
    \caption{Color images composed of S-2 and WV-2 bands to inspect the co-registration correctness of the prepared data. The example shows correct co-registration among (a) S-2 and (b) WV-2 bands, (c) between S-2 and WV-2 images (in a form of a checkerboard image), and  (d)~among multiple S-2 images (incorrect co-registration would result in color artefacts).}
    \label{fig:color_example}
\end{figure}

Overall, we acquired 91 pairs of HR images, each of which is coupled with multiple LR images (every HR and LR image is composed of four bands). To verify whether the bands are correctly co-registered after cropping, we composed color images and inspected them against color artefacts (Fig.~\ref{fig:color_example}a,\,b). In order to check the co-registration correctness between LR images and the HR image, we have assembled checkerboard mosaicked images and inspected them visually for all the scenes (Fig.~\ref{fig:color_example}c). Finally, we combined multiple S-2 images for each scene to inspect whether the obtained images are free of color and halo artefacts that could result from poor co-registration across multiple S-2 images and/or the spectral bands (Fig.~\ref{fig:color_example}d).

\subsection*{The Evaluation Procedure}
 The evaluated SR process is fed with $N$ input images for reconstructing a specific S-2 band to produce a single super-resolved image ($\srout$) which is subsequently compared with the corresponding $\hrref$. By an input image we understand a single S-2 band, but it is also possible that multiple bands cropped to show the same area are processed to reconstruct a single S-2 band or multiple bands, and each of the super-resolved 10\,m bands can be compared with a corresponding $\hrref$ derived from the WV-2 data. We measure the similarity between $\srout$ and $\hrref$ with the PSNR, SSIM, and LPIPS metrics that are commonly employed for assessing the SR quality~\cite{WangBayram2022}. ~$\lrinput$ and $\hrref$ are acquired by different sensors, hence the differences between $\srout$ and $\hrref$ result not only from the lack of reconstruction accuracy, but also from the characteristics of the imaging sensor and the temporal changes of the scene (due to different acquisition time)~\cite{Benecki2018AA}. Also, as the $N$ input images are co-registered in the LR space at whole-pixel precision, the super-resolved $\srout$ may be displaced up to $\alpha$ pixels in each dimension compared with $\hrref$. For Proba-V images (acquired with the same satellite), it was sufficient to co-register $\srout$ to $\hrref$ and to compensate the brightness bias~\cite{Martens2019}. The co-registration was performed by shifting $\srout$ by $\left[-\alpha;\alpha\right]$ pixels in vertical and horizontal dimensions to maximize the PSNR score computed for the whole $\hrref$ image without an $\alpha$-wide boundary (so that even after shifting the same part of $\hrref$ is considered for computing the score). For \benchmark, we have fully adopted that approach. In~\cite{Martens2019}, the brightness was compensated by subtracting the mean brightness from both $\srout$ and $\hrref$. In the case considered here, the differences between S-2 and WV-2 images are more substantial than among Proba-V images, therefore we match the histogram of $\srout$ to $\hrref$ before the evaluation. Importantly, this does not convey any HR information from $\hrref$ to $\srout$, but only compensates the low-frequency differences between them. Also, we do not modify $\hrref$, so that every $\srout$ obtained using different techniques is always compared to the same reference image.

As proposed in~\cite{Martens2019}, in addition to reporting the direct values of PSNR, we also compute their relation to the scores retrieved for images obtained by averaging LR images enlarged with bicubic interpolation, treated as a baseline which does not offer any information gain. We adopt the same approach for SSIM and LPIPS metrics, and we also compute the balanced metric as:
\begin{equation}
    \mathcal{B}(\srout)=\frac{1}{3} \left[ \frac{{\rm PSNR}\left(\bicout,\hrref\right)}{{\rm PSNR}\left(\srout,\hrref\right)} + \frac{ {\rm SSIM}\left(\bicout, \hrref\right)}{ {\rm SSIM}\left(\srout,\hrref\right)} + \frac{ {\rm LPIPS}\left(\srout,\hrref\right)}{ {\rm LPIPS}\left(\bicout,\hrref\right)} \right] ,
\end{equation}
where 
$\bicout$ is obtained by averaging all bicubically-upsampled $\lrinput$'s in the scene (they are all co-registered as justified earlier). Hence, $\mathcal{B}<1$ means better performance compared with the bicubic interpolation, and $\mathcal{B}>1$ indicates the opposite case (for PSNR and SSIM the higher score indicates higher similarity, and for LPIPS the lower the score, the higher the similarity).

\begin{figure}[!htb]
\centering
\newcommand{\mywidth}{0.19}
\renewcommand{\tabcolsep}{0.3mm}
\begin{tabular}{ccccc}
\includegraphics[width=\mywidth\textwidth]{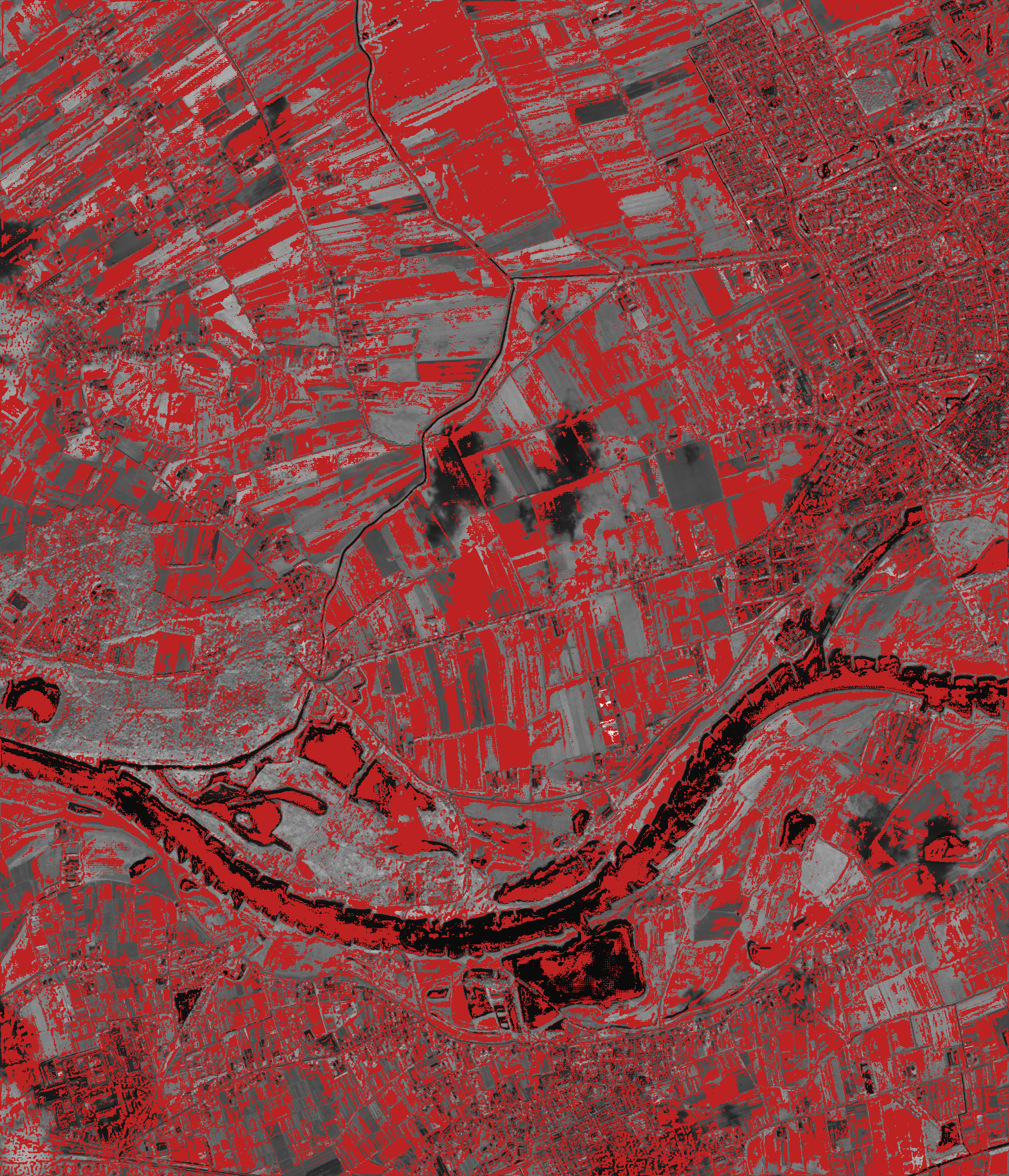} &
\includegraphics[width=\mywidth\textwidth]{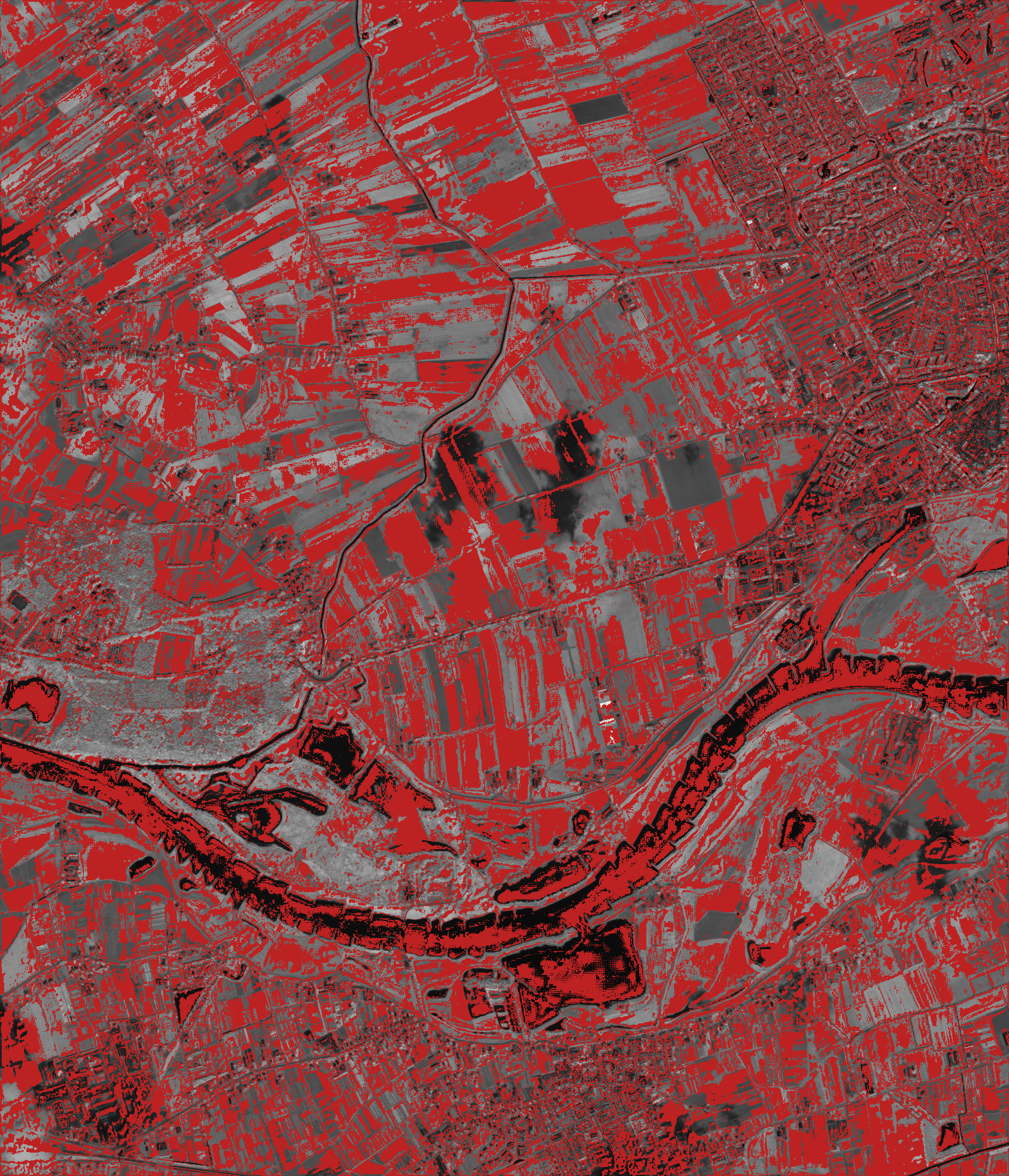} &
\includegraphics[width=\mywidth\textwidth]{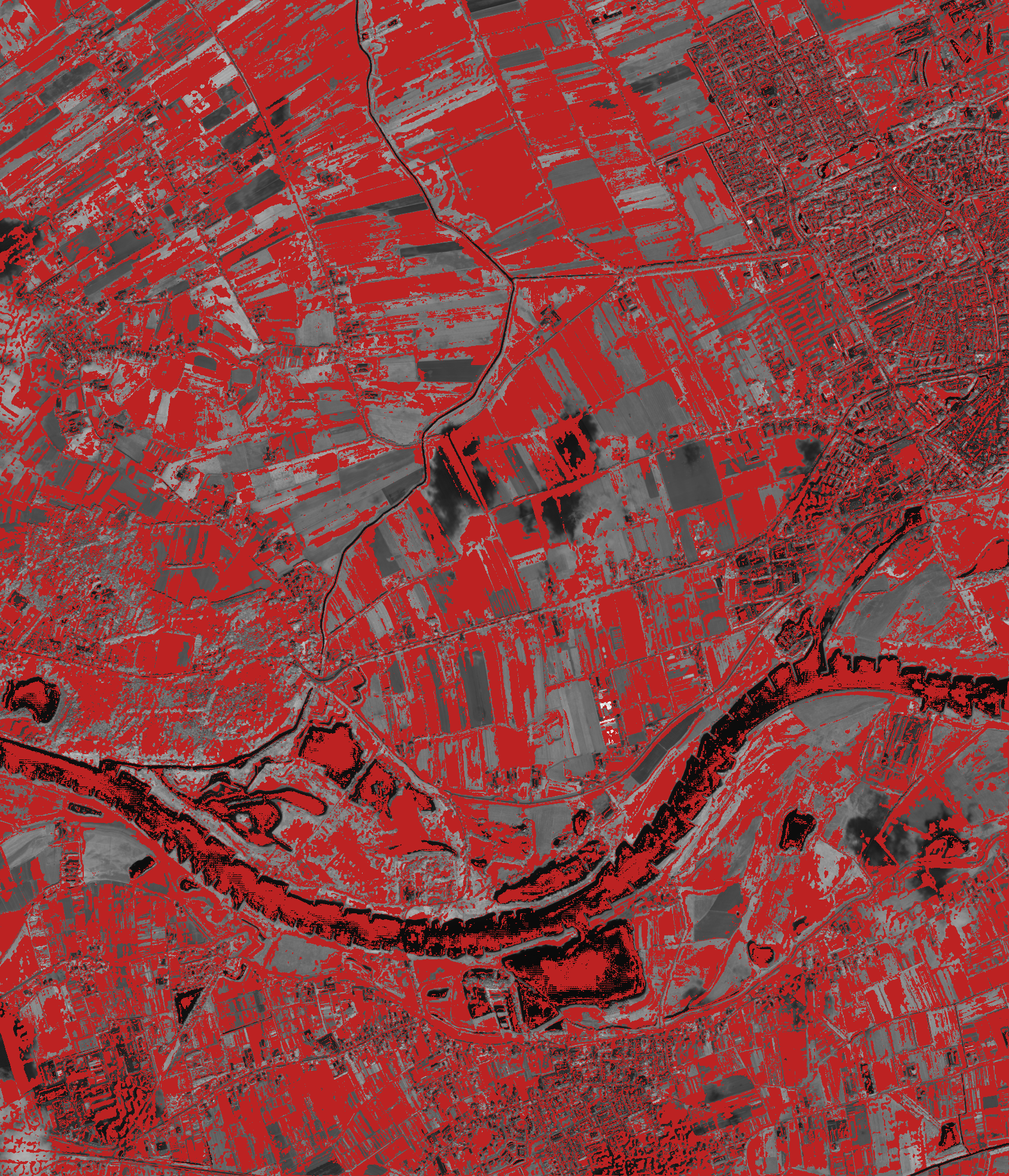} &
\includegraphics[width=\mywidth\textwidth]{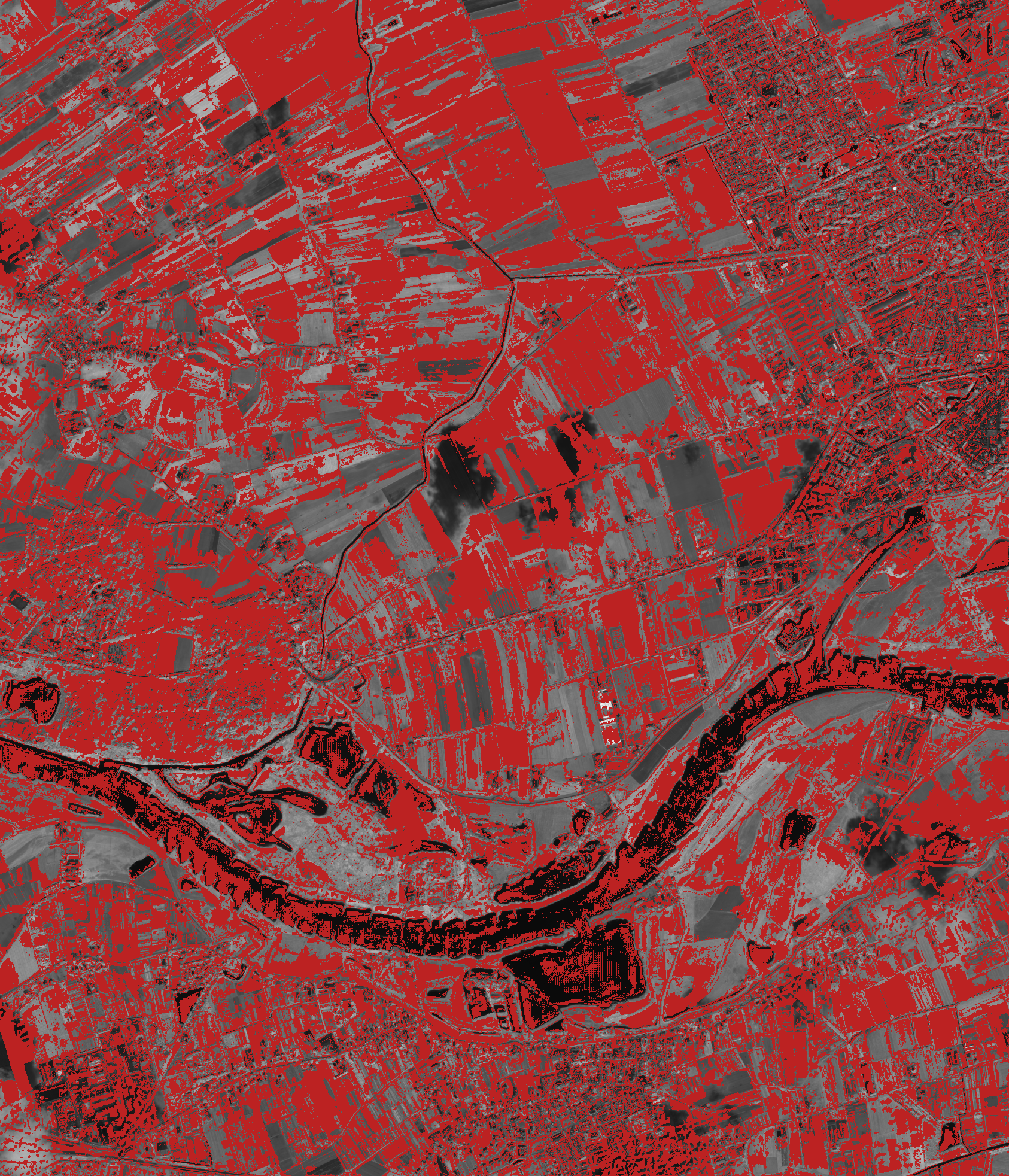} &
\includegraphics[width=\mywidth\textwidth]{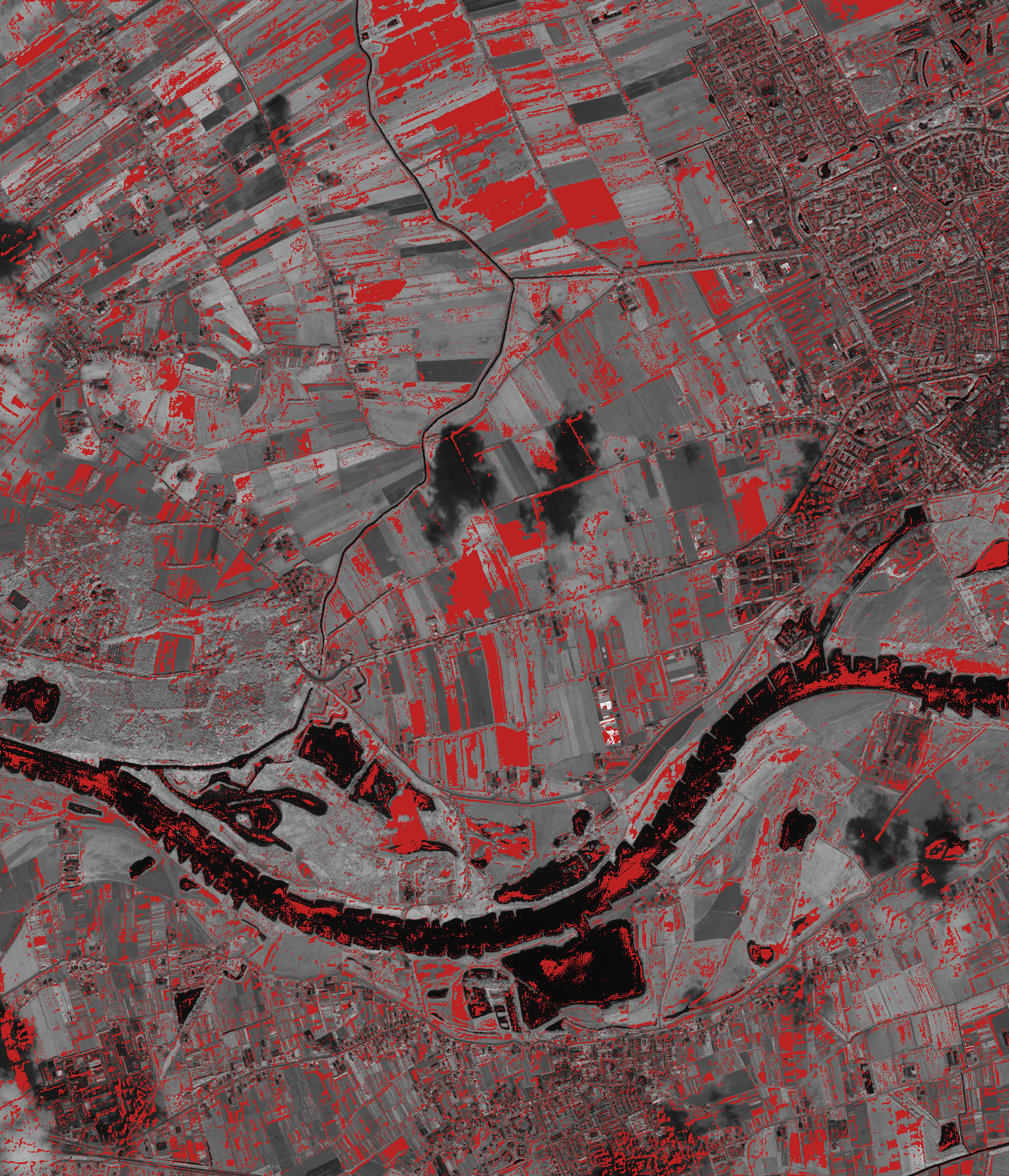} \\
(a) & (b) & (c) & (d) & (e) \\
\end{tabular}
\caption{Areas (red) in which bicubically interpolated image is closer to the HR reference than the super-resolved one (after applying appropriate shift and brightness compensation). Outcomes of (a, b)~HighRes-net~\cite{deudon2020highresnet} and (c, d)~RAMS~\cite{Salvetti2020} models trained over Proba-V NIR and Red images, respectively, are shown along with (e)~the final relevance mask (red areas are masked out).}
\label{fig:differences}
\end{figure}

During our experiments (reported later in this paper), we observed that the PSNR and SSIM metrics are not particularly effective in assessing the reconstruction accuracy when $\lrinput$ and $\hrref$ originate from different satellites---the scores differ little across $\srout$'s obtained with the employed SR or interpolation procedures. Based on experimental validation and MOS survey, whose results are discussed later in this paper, it is clear that the outcomes obtained using MISR techniques better reflect the image details than relying on bicubic interpolation, and this is correctly captured relying on the LPIPS metric. It can be seen from Figure~\ref{fig:differences} that there are regions, marked as red in Figure~\ref{fig:differences}e, in which bicubic interpolation consistently prevails over all of the four considered SR techniques trained from real-life data in terms of the mean squared error (MSE), which the PSNR metric is based on (these regions were extracted after applying shift and brightness compensation). This may result from the temporal differences between $\lrinput$'s and $\hrref$---in the regions where bicubic interpolation renders lower MSE than all the SR techniques, the HR reference can be regarded as inappropriate to assess the reconstruction quality. It can be noticed that such regions are mainly located in the cultivated fields, while in the highly-detailed urban areas they are isolated (this may result from shadows and differences in lighting conditions). Overall, for every scene, we generate the relevance masks which indicate the regions in $\hrref$ where at least one MISR model (out of four considered in our study) offers better performance than the bicubic interpolation in terms of MSE, and if the relevance masks are applied, then the metrics are computed only inside such regions (hence, in Figure~\ref{fig:differences}e, the pixels marked as red would be ignored when computing the metrics with the relevance masks applied).

\section*{Data Records}

\begin{figure}
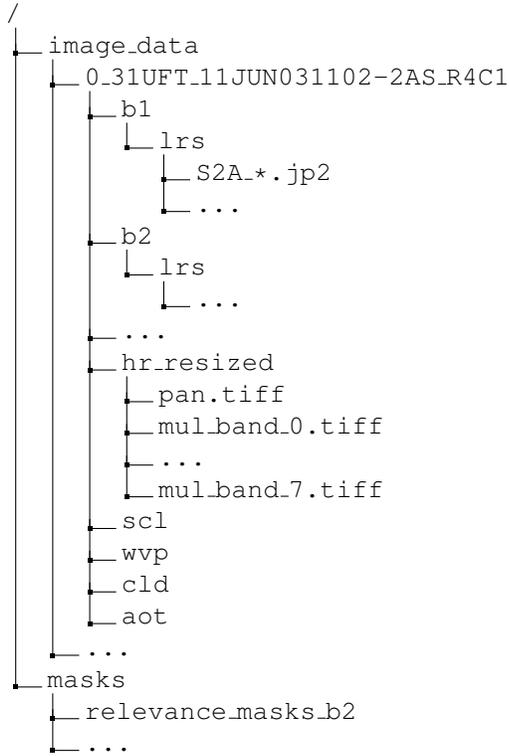

\dirtree{%
.1 /.
.2 image\_data.
.3 0\_31UFT\_11JUN031102-2AS\_R4C1.
.4 b1.
.5 lrs.
.6 S2A\_*.jp2.
.6 ....
.4 b2.
.5 lrs.
.6 ....
.4 ....
.4 hr\_resized.
.5 pan.tiff.
.5 mul\_band\_0.tiff.
.5 ....
.5 mul\_band\_7.tiff.
.4 scl.
.4 wvp.
.4 cld.
.4 aot.
.3 ....
.2 masks.
.3 relevance\_masks\_b2.
.3 ....
}

    \caption{The folder structure of the MuS2 benchmark dataset.}
    \label{fig:files}
\end{figure}

The MuS2 benchmark dataset is available at \url{https://doi.org/10.7910/DVN/1JMRAT} and its folder structure is depicted in Figure~\ref{fig:files}. The images are grouped into 91 scenes, each of which is placed in a separate scene-level folder, whose names follow the pattern {\tt <ord\_num>\_<mgrs>\_<date>-2AS\_<tile>}, where {\tt <ord\_num>} is the ordinal number of Sentinel-2 product, {\tt <mgrs>} is the MGRS tile representing the captured area, {\tt <date>} is the acquisition date of the WV-2 image, and {\tt <tile>} is the name of the WV-2 tile. Each scene-level folder contains 12 band-level folders ({\tt b1}, {\tt b2}, {\tt b3}, etc.) with the corresponding S-2 bands. Every band-level folder contains 14 or 15 Level-2A images captured at different revisits. The WV-2 bands (numbered from 0 to 7) along with a panchromatic image are stored in the {\tt hr\_resized} folder for each scene, resized to the dimensions $3\times$ larger than the S-2 input images. Also, the scene classification maps ({\tt scl}), cloud masks ({\tt cld}), aerosol optical thickness ({\tt aot}), and water vapour maps ({\tt wvp}) are included as well. The resolution of HR images is $3\times$ larger than the resolution of S-2 10\,m GSD images (i.e., bands B02, B03, B04, and B08). S-2 and WV-2 band images are intended to be coupled for evaluation as follows:
\begin{itemize}
    \item S-2 B02 ({\tt b2}) with WV-2 B band ({\tt mul\_band\_1}),
    \item S-2 B03 ({\tt b3}) with WV-2 G band ({\tt mul\_band\_2}),
    \item S-2 B04 ({\tt b4}) with WV-2 R band ({\tt mul\_band\_4}),
    \item S-2 B08 ({\tt b8}) with WV-2 NIR1 band ({\tt mul\_band\_6}).
\end{itemize}
Additionally, the relevance masks for the 10\,m S-2 bands are also included in the {\tt masks} folder.

\section*{Technical Validation} \label{sec:exp}

The goal of our experiments was twofold: (\textit{i})~to confirm that \benchmark, including the evaluation procedure, is suitable for assessing the reconstruction accuracy, and (\textit{ii})~to report the scores of the well-established SR techniques, which can be used as a baseline for future research. For this sake, we included HighRes-net that recursively combines latent LR representations to obtain the super-resolved image~\cite{deudon2020highresnet}, as well as the residual attention MISR (RAMS)  network equipped with the attention mechanism~\cite{Salvetti2020}.

In this study, we focus on evaluating MISR performed for four 10\,m S-2 bands matched with the HR reference obtained from WV-2. In order to verify that based on \benchmark, we can assess how much the reconstructed details resemble those observed in $\hrref$, we consider three groups of methods: (\textit{i})~MISR networks trained with real-world LR and HR Proba-V images~\cite{Martens2019}, (\textit{ii})~MISR networks trained using S-2 simulated LR data~\cite{Kawulok2019IGARSS} (obtained by degrading an original S-2 image, later treated as $\hrref$), and (\textit{iii})~image interpolation techniques treated as a baseline. In~\cite{Kawulok2021IGARSS}, we demonstrated that RAMS and HighRes-net trained from Proba-V images are successful in super-resolving S-2 data, and they do not generate artefacts observed when these networks are trained using simulated LR images. To verify how such artefacts influence the scores for \benchmark, we report the performance for the networks trained with Proba-V NIR, Proba-V Red, and S-2 simulated images.

\newcommand{\nn}{NN}
\newcommand{\bicubic}{Bicubic}
\newcommand{\lanczos}{Lanczos}
\newcommand{\linear}{Linear}
\newcommand{\hrnir}{Pr-NIR}
\newcommand{\hrred}{Pr-Red}
\newcommand{\hrsim}{S2-simul.}
\newcommand{\ramsnir}{Pr-NIR}
\newcommand{\ramsred}{Pr-Red}
\newcommand{\ramssim}{S2-simul.}

\begin{table*}[!t]
\renewcommand{\tabcolsep}{1mm}
    \centering
        \caption{Reconstruction accuracy obtained for our \benchmark\, benchmark measured with PSNR (in dB), SSIM, LPIPS, and the balanced score $\mathcal{B}$, obtained for different interpolation techniques alongside HighRes-net and RAMS networks trained using real-world and simulated images. The scores are reported without and with the relevance masks. The best scores (highest PSNR and SSIM, and lowest LPIPS and $\mathcal{B}$) are boldfaced.}

\resizebox{\textwidth}{!}{
    \begin{tabular}{llccccccccccccccccccc}

\Xhline{3\arrayrulewidth}

& & \multicolumn{4}{c}{Band B02} & &\multicolumn{4}{c}{Band B03} & & \multicolumn{4}{c}{Band B04} & & \multicolumn{4}{c}{Band B08} \\ \cline{3-6} \cline{8-11} \cline{13-16} \cline{18-21}

\multicolumn{2}{r}{Method $\downarrow$} & PSNR & SSIM & LPIPS & $\mathcal{B}$ & & PSNR & SSIM & LPIPS & $\mathcal{B}$ & & PSNR & SSIM & LPIPS & $\mathcal{B}$ & & PSNR & SSIM & LPIPS & $\mathcal{B}$ \\ \Xhline{3\arrayrulewidth}

\multicolumn{9}{l}{{No mask applied (whole image compared):}} \\ \cline{2-21}

\multirow{4}{*}{\begin{sideways} Interp. \end{sideways}}  & \nn & $	34.85$ & $	.9266$ & $	.2705$ & $	0.984$ & 	& $	30.66$ & $	.8465$ & $	.3733$ & $	0.979$ & 	& $	31.80$ & $	.8575$ & $	.3451$ & $	0.988$ & 	& $	24.06$ & $	.5694$ & $	.5285$ & $	1.001$	\\
& \linear & $	34.92$ & $	\bf{.9305}$ & $	.2986$ & $	1.016$ & 	& $	30.72$ & $	.8533$ & $	.4157$ & $	1.012$ & 	& $	31.85$ & $	.8627$ & $	.3762$ & $	1.015$ & 	& $	24.10$ & $	.5790$ & $	.5318$ & $	0.998$	\\
& \bicubic & $	34.95$ & $	.9301$ & $	.2858$ & $	1.000$ & 	& $	30.75$ & $	.8535$ & $	.4025$ & $	1.000$ & 	& $	31.88$ & $	.8631$ & $	.3609$ & $	1.000$ & 	& $	24.14$ & $	.5835$ & $	.5419$ & $	1.000$	\\
& \lanczos & $	34.95$ & $	.9302$ & $	.2817$ & $	0.995$ & 	& $	30.76$ & $	\bf{.8540}$ & $	.3974$ & $	0.996$ & 	& $	\bf{31.89}$ & $	\bf{.8636}$ & $	.3568$ & $	0.996$ & 	& $	\bf{24.15}$ & $	.5853$ & $	.5408$ & $	0.998$	\\ \cline{2-21}
\multirow{3}{*}{\begin{sideways} HRn \end{sideways}}  & \hrnir & $	\bf{35.19}$ & $	.9288$ & $	.2310$ & $	0.936$ & 	& $	\bf{30.84}$ & $	.8498$ & $	.3214$ & $	0.934$ & 	& $	31.82$ & $	.8604$ & $	.2864$ & $	0.932$ & 	& $	23.93$ & $	.5820$ & $	\bf{.4330}$ & $	\bf{0.936}$	\\
& \hrred & $	34.95$ & $	.9229$ & $	.2192$ & $	0.925$ & 	& $	30.68$ & $	.8437$ & $	.3164$ & $	0.933$ & 	& $	31.61$ & $	.8553$ & $	.2792$ & $	0.929$ & 	& $	23.98$ & $	.5842$ & $	.4446$ & $	0.941$	\\
& \hrsim & $	34.63$ & $	.9125$ & $	.2730$ & $	0.998$ & 	& $	30.39$ & $	.8247$ & $	.3610$ & $	0.988$ & 	& $	31.48$ & $	.8383$ & $	.3412$ & $	1.004$ & 	& $	23.69$ & $	.5311$ & $	.4901$ & $	1.014$	\\ \cline{2-21}
\multirow{3}{*}{\begin{sideways} RAMS \end{sideways}}  & \ramsnir & $	35.01$ & $	.9232$ & $	.2166$ & $	0.922$ & 	& $	30.61$ & $	.8400$ & $	.3171$ & $	0.936$ & 	& $	31.59$ & $	.8543$ & $	.2757$ & $	0.927$ & 	& $	23.97$ & $	.5850$ & $	.4391$ & $	0.937$	\\
& \ramsred & $	35.11$ & $	.9257$ & $	\bf{.2165}$ & $	\bf{0.917}$ & 	& $	30.80$ & $	.8478$ & $	\bf{.3107}$ & $	\bf{0.925}$ & 	& $	31.87$ & $	.8605$ & $	\bf{.2706}$ & $	\bf{0.915}$ & 	& $	24.01$ & $	\bf{.5854}$ & $	.4558$ & $	0.947$	\\
& \ramssim & $	34.55$ & $	.9143$ & $	.2580$ & $	0.984$ & 	& $	30.35$ & $	.8238$ & $	.3660$ & $	0.990$ & 	& $	31.51$ & $	.8345$ & $	.3503$ & $	1.006$ & 	& $	23.84$ & $	.5261$ & $	.5617$ & $	1.056$	\\ \hline
\multicolumn{9}{l}{{Relevance masks applied:}} \\ \cline{2-21} \multirow{4}{*}{\begin{sideways} Interp. \end{sideways}}  & \nn & $	34.60$ & $	.9613$ & $	.2226$ & $	0.921$ & 	& $	30.39$ & $	.9359$ & $	.3021$ & $	0.891$ & 	& $	31.56$ & $	.9091$ & $	.2710$ & $	0.902$ & 	& $	24.01$ & $	.8327$ & $	.4149$ & $	0.821$	\\
& \linear & $	34.64$ & $	.9628$ & $	.2295$ & $	0.928$ & 	& $	30.42$ & $	.9381$ & $	.3132$ & $	0.899$ & 	& $	31.58$ & $	.9114$ & $	.2810$ & $	0.911$ & 	& $	24.03$ & $	.8360$ & $	.4239$ & $	0.825$	\\
& \bicubic & $	34.70$ & $	.9626$ & $	.2225$ & $	0.920$ & 	& $	30.48$ & $	.9383$ & $	.3049$ & $	0.892$ & 	& $	31.65$ & $	.9119$ & $	.2716$ & $	0.901$ & 	& $	24.11$ & $	.8385$ & $	.4156$ & $	0.818$	\\
& \lanczos & $	34.71$ & $	.9627$ & $	.2203$ & $	0.917$ & 	& $	30.49$ & $	\bf{.9384}$ & $	.3023$ & $	0.889$ & 	& $	31.66$ & $	.9122$ & $	.2694$ & $	0.899$ & 	& $	24.12$ & $	.8392$ & $	.4119$ & $	0.815$	\\ \cline{2-21}
\multirow{3}{*}{\begin{sideways} HRn \end{sideways}}  & \hrnir & $	\bf{35.41}$ & $	\bf{.9630}$ & $	.2021$ & $	0.892$ & 	& $	\bf{31.06}$ & $	\bf{.9384}$ & $	.2693$ & $	0.858$ & 	& $	\bf{32.14}$ & $	.9135$ & $	.2404$ & $	0.866$ & 	& $	24.40$ & $	.8401$ & $	.3503$ & $	0.774$	\\
& \hrred & $	35.24$ & $	.9608$ & $	.1987$ & $	0.889$ & 	& $	30.96$ & $	.9369$ & $	.2667$ & $	0.855$ & 	& $	31.96$ & $	.9111$ & $	.2375$ & $	0.865$ & 	& $	24.33$ & $	.8404$ & $	.3626$ & $	0.782$	\\
& \hrsim & $	34.57$ & $	.9555$ & $	.2425$ & $	0.948$ & 	& $	30.32$ & $	.9272$ & $	.3210$ & $	0.912$ & 	& $	31.45$ & $	.8983$ & $	.2949$ & $	0.934$ & 	& $	23.81$ & $	.8152$ & $	.4271$ & $	0.837$	\\ \cline{2-21}
\multirow{3}{*}{\begin{sideways} RAMS \end{sideways}}  & \ramsnir & $	35.26$ & $	.9611$ & $	\bf{.1939}$ & $	\bf{0.882}$ & 	& $	30.88$ & $	.9352$ & $	.2675$ & $	0.857$ & 	& $	31.89$ & $	.9108$ & $	.2338$ & $	0.863$ & 	& $	\bf{24.51}$ & $	.8406$ & $	\bf{.3394}$ & $	\bf{0.765}$	\\
& \ramsred & $	35.29$ & $	.9619$ & $	.1953$ & $	\bf{0.882}$ & 	& $	30.98$ & $	.9379$ & $	\bf{.2621}$ & $	\bf{0.851}$ & 	& $	32.11$ & $	\bf{.9144}$ & $	\bf{.2329}$ & $	\bf{0.858}$ & 	& $	24.47$ & $	\bf{.8411}$ & $	.3567$ & $	0.776$	\\
& \ramssim & $	34.39$ & $	.9544$ & $	.2393$ & $	0.950$ & 	& $	30.19$ & $	.9241$ & $	.3315$ & $	0.923$ & 	& $	31.41$ & $	.8951$ & $	.3090$ & $	0.944$ & 	& $	23.97$ & $	.8122$ & $	.4721$ & $	0.863$	\\

\Xhline{3\arrayrulewidth}

\multicolumn{20}{l}{HRn---HighRes-net, trained with Proba-V NIR (Pr-NIR), Proba-V Red (Pr-Red), or S-2 simulated (Simul.) images} \\

\multicolumn{20}{l}{NN---nearest-neighbor interpolation}

    \end{tabular}
    }

    \label{tab:results}
\end{table*}
\begin{figure*}[!h]
\centering
\includegraphics[width=\textwidth]{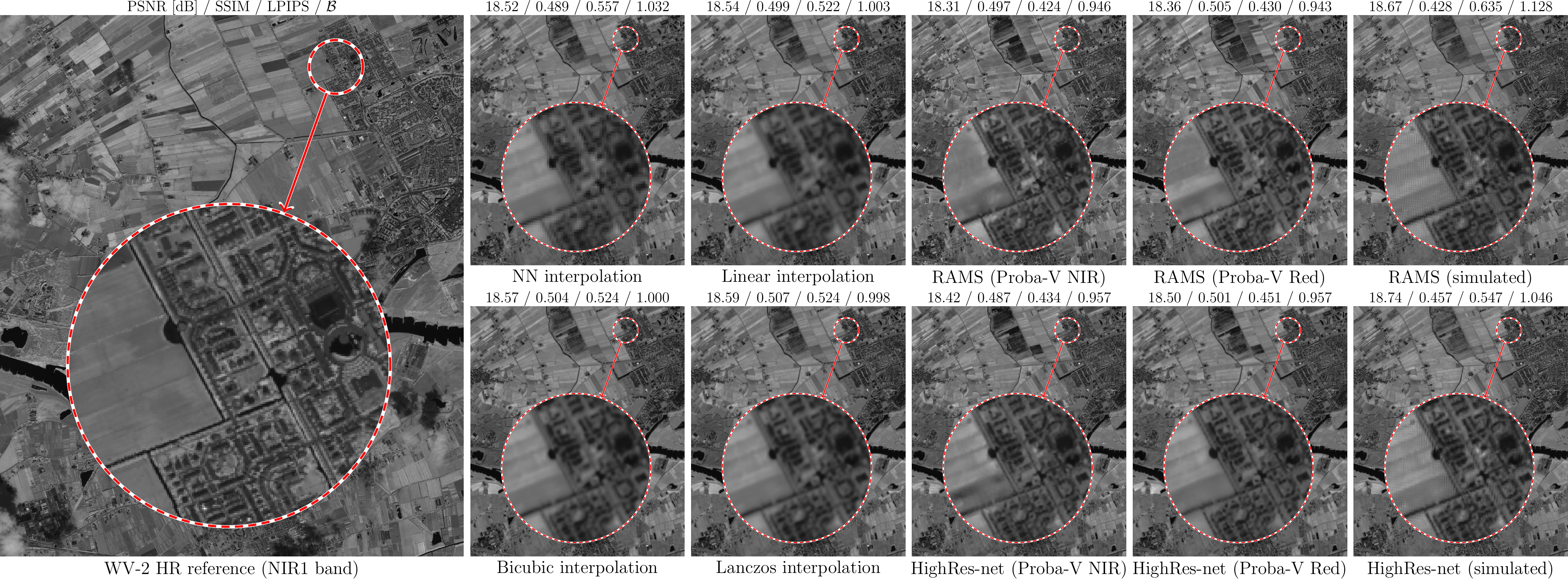}
\caption{Reconstruction outcome (band B08) obtained with RAMS and HighRes-net trained from real-life Proba-V NIR and Red images and from simulated data, compared with image interpolation techniques and HR reference. PSNR (in dB), SSIM, LPIPS and $\mathcal{B}$ scores are presented above each example.}
\label{fig:results}
\end{figure*}

Quantitative results obtained for four S-2 bands without and with the relevance masks applied are reported in Table~\ref{tab:results}. It can be seen that without the relevance mask, the PSNR and SSIM scores do not differ much between each other within particular bands, and they do not indicate the SR techniques to be better than interpolation. This is against what can be judged from visual inspection of the reconstruction outcome, an example of which is presented in Figure~\ref{fig:results}. The images rendered by HighRes-net and RAMS trained over Proba-V images present more details than the interpolated images (which are quite blurry), and they are free from the artefacts present for the models trained from the S-2 simulated data. It is worth noting that based on visual assessment, the details in $\srout$ are quite close to those observed in the WV-2 image, which suggests that they were reconstructed indeed rather than hallucinated (such hallucination artefacts are quite common for SISR techniques, especially for larger magnification factors~\cite{Ledig2017}). Unfortunately, these qualitative observations are quantitatively reflected only in the LPIPS values, while PSNR and SSIM are slightly worse for HighRes-net and RAMS---overall, both LPIPS and $\mathcal{B}$ indicate that SR networks perform better than interpolation and they penalize for the artefacts. Apparently, the differences between $\lrinput$ and $\hrref$ images resulting from different image acquisition conditions prevail over the accuracy of reconstructing the details when local pixel-wise metrics like PSNR or SSIM are used, but they have smaller impact on the feature-based LPIPS metric. When the relevance masks are applied, all the metrics indicate the superiority of SR networks and their $\mathcal{B}$ scores are aligned in a similar order as without the mask.

As shown in Table~\ref{tab:results}, the reported metrics (without the relevance masks) are not consistent in indicating the SR performance. In order to verify which of them is most reliable for assessing the reconstruction accuracy, we have prepared a MOS survey composed of 15 queries\footnote{The MOS questions are presented in the Supplementary Material.}. We considered the cases in which PSNR, SSIM, and LPIPS metrics pointed to a different SR or interpolation outcome as the most (eight cases) or the least (seven cases) similar to $\hrref$. In each query, the participants of diverse background (including Earth observation professionals and individuals without any remote sensing experience) were presented such three images alongside $\hrref$, and asked to select the best or the worst image for retrieving some specific details (e.g., delineating the roads). In this way, we wanted to prevent the participants from being biased toward picking an image of the best perceptual quality, instead of the one whose details are most similar to those in $\hrref$. The results of the survey (averaged from over 160 responses) are reported in Table~\ref{tab:mos}---clearly, LPIPS correlates most with the human judgement both in picking the most or least accurate reconstruction outcome (retrieved with an SR network trained from Proba-V data and an interpolation technique, respectively). Overall, even though LPIPS is a perceptual metric, the reported study shows that it can be exploited as a reliable indicator of reconstruction accuracy in this case. We expect that the balanced metric $\mathcal{B}$ introduces an additional stability, as the PSNR and SSIM scores are sensitive to the hallucination effects~\cite{Ledig2017} which potentially may not affect the LPIPS metric.
\begin{table}[h]
    \centering
    \caption{The MOS survey outcome showing how often each metric was consistent with the answers  (in \%), stratified into SR nets trained with Proba-V and simulated images, and interpolation.}
    \begin{tabular}{lrrrrr}
    \Xhline{2\arrayrulewidth}
    \multicolumn{2}{l}{$\downarrow$ Query type~~~~~~~~Method $\downarrow$} & \multicolumn{1}{c}{PSNR} & \multicolumn{1}{c}{SSIM} & \multicolumn{1}{c}{LPIPS} & \multicolumn{1}{c}{Total} \\ \Xhline{2\arrayrulewidth}
    \multirow{4}{*}{\begin{sideways}Best acc.\end{sideways}}
    & SR networks (Proba-V) & 	$3.92$ & 	$14.46$ & 	$55.57$ & 	$73.95$ \\
    & SR networks (simulated) & 	$11.67$ & 	$0.00$ & 	$7.61$ & 	$19.28$ \\
    & Image interpolation & 	$1.81$ & 	$4.97$ & 	$0.00$ & 	$6.78$ \\ \cline{2-6}
    & Total & 	$17.39$ & 	$19.43$ & 	$63.18$ & 	$100.00$ \\
    \hline
    \multirow{4}{*}{\begin{sideways}Worst acc.\end{sideways}}
    & SR networks (Proba-V) & 	$5.85$ & 	$0.00$ & 	$0.00$ & 	$5.85$ \\
    & SR networks (simulated) & 	$2.07$ & 	$10.93$ & 	$6.71$ & 	$19.71$ \\
    & Image interpolation & 	$0.00$ & 	$0.00$ & 	$74.44$ & 	$74.44$ \\ \cline{2-6}
    & Total & 	$7.92$ & 	$10.93$ & 	$81.15$ & 	$100.00$ \\
 	  \Xhline{2\arrayrulewidth}

    \end{tabular}
    \label{tab:mos}
\end{table}


\subsection*{Conclusions and outlook}
In this paper, we introduced a new \benchmark\, benchmark for assessing the accuracy of MISR for S-2 images. It is composed of original S-2 data coupled with HR references obtained from WV-2. We have proposed the evaluation protocol based on a balanced score built upon PSNR, SSIM, and LPIPS metrics, and we demonstrated that it is suitable for assessing the reconstruction accuracy. Based on the MOS survey, we showed that LPIPS metric can be employed for assessing the similarity to the ground truth and that it is robust against variations resulting from acquiring images by different satellites. Additionally, we have elaborated the relevance masks which pick the regions to allow for pixel-wise evaluation performed with ``traditional'' PSNR and SSIM metrics.

Although we have shown that all of the S-2 10\,m bands can be coupled with corresponding WV-2 bands, evaluating reconstruction of the remaining bands, potentially relying on panchromatic WV-2 images, remains an open issue. With our data preparation procedure, it is possible to crop all the bands which may help address this challenging problem in the future. Furthermore, here we consider the magnification factor of $3\times$, but with the prepared procedure, other factors may be considered as well.
Finally, even though the reconstruction outcomes obtained with the reported techniques are of definitely higher quality than the bicubically-interpolated images, they are still far from the HR reference. We expect that \benchmark\, used as a test set for assessing emerging SR techniques will guide the researchers toward developing more effective SR techniques. They may result from improved architectures, better data simulation and augmentation procedures, as well as from new loss functions that would allow for more robust training from real-world data.

\section*{Usage Notes}

After downloading the dataset from \url{https://doi.org/10.7910/DVN/1JMRAT}, the SR algorithms can be fed with multiple images for each scene (and for each band). Every S-2 band can be reconstructed relying on multiple images of the same band, but other bands can also be exploited during reconstruction. Also, the data can be used for SISR by picking a single LR input image for each scene and band.

To assess the quality of the super-resolved images, the code published at the Code Ocean (\url{https://codeocean.com/capsule/8131193/tree/v1}) can be exploited to compute the metrics (PSNR, SSIM, LPIPS, and the balanced score $\mathcal{B}$).

\section*{Code availability}

The code for creating the benchmark from raw data and for evaluating the SR outcome is available at \url{https://codeocean.com/capsule/8131193/tree/v1}. The code is documented and accompanied with usage examples.

\section*{Acknowledgements}

The work was funded by European Space Agency (PIGEON project). This research was supported by the National Science Centre, Poland, under Research Grant 2019/35/B/ST6/03006 (TT, JN, MK). MK was supported by the SUT funds through the Rector’s Research and Development Grant 02/080/RGJ22/0024.

\section*{Author contributions statement}

\textbf{Pawel Kowaleczko}: Software, Validation, Investigation, Data Curation, Writing---Original Draft, Visualization;
\textbf{Tomasz Tarasiewicz}: Software, Data Curation;
\textbf{Maciej~Ziaja}: Software, Validation, Data Curation, Visualization;
\textbf{Daniel~Kostrzewa}: Methodology, Validation, Investigation, Resources, Writing---Original Draft;
\textbf{Jakub~Nalepa}: Methodology, Validation, Writing---Review \& Editing;
\textbf{Przemyslaw~Rokita}: Validation, Writing---Review \& Editing;
\textbf{Michal~Kawulok}: Conceptualization, Methodology, Validation, Writing---Original Draft, Writing---Review \& Editing, Visualization, Supervision, Project administration, Funding acquisition;
All authors reviewed the manuscript.

\section*{Competing interests}
The authors declare that they have no known competing financial interests or personal relationships that could have appeared to influence the work reported in this paper.

\newcommand{\supplementfilename}{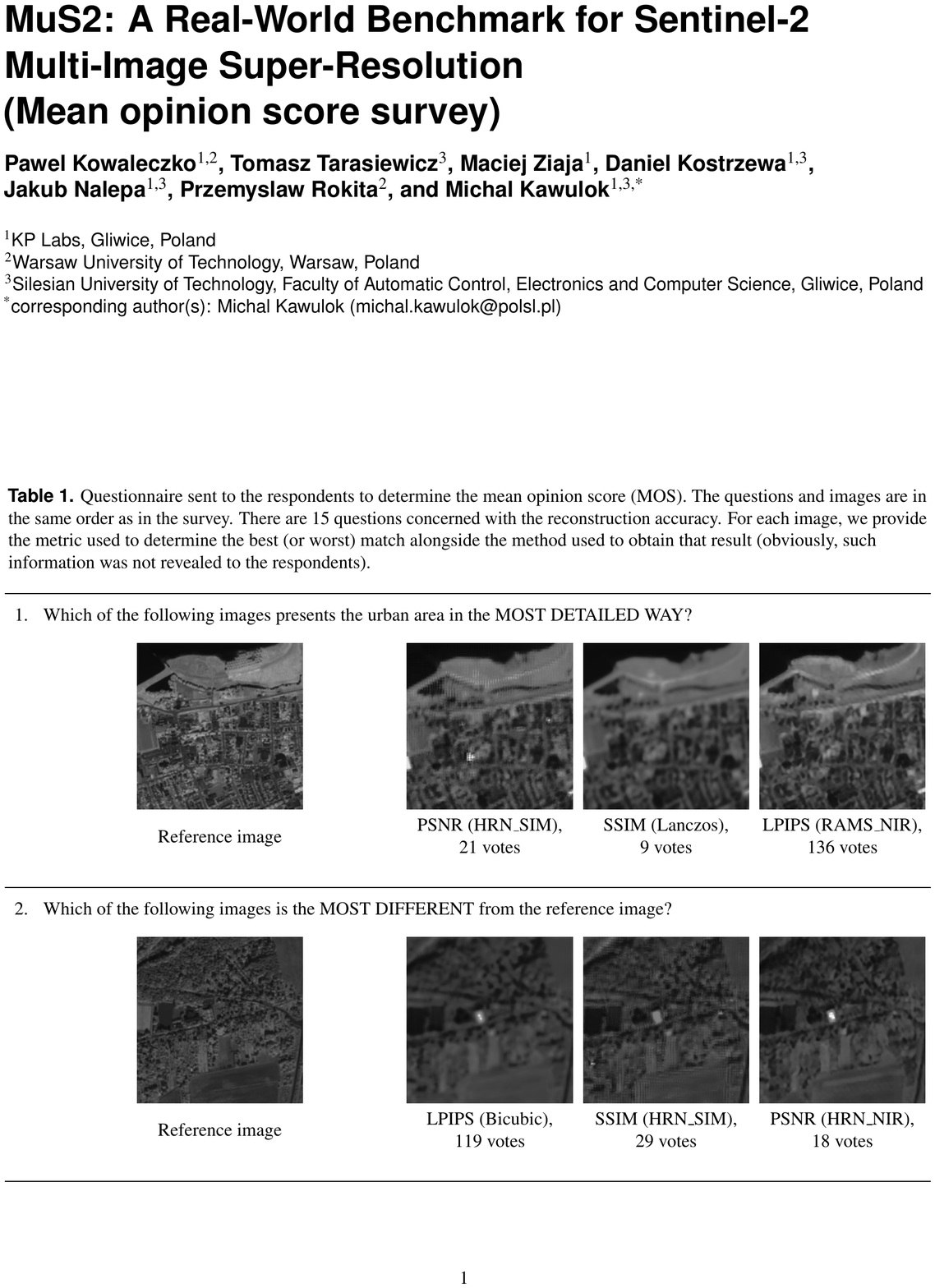}

\pdfximage{\supplementfilename}
\def\numbersupplementpages{\the\pdflastximagepages}




\section*{Supplementary material}

# MuS2: A Real-World Benchmark for Sentinel-2 Multi-Image Super-Resolution (Mean opinion score survey)

**Pawel Kowaleczko**[1,2], **Tomasz Tarasiewicz**[3], **Maciej Ziaja**[1], **Daniel Kostrzewa**[1,3], **Jakub Nalepa**[1,3], **Przemyslaw Rokita**[2], **and Michal Kawulok**[1,3,*]

[1]KP Labs, Gliwice, Poland
[2]Warsaw University of Technology, Warsaw, Poland
[3]Silesian University of Technology, Faculty of Automatic Control, Electronics and Computer Science, Gliwice, Poland
[*]corresponding author(s): Michal Kawulok (michal.kawulok@polsl.pl)

**Table 1.** Questionnaire sent to the respondents to determine the mean opinion score (MOS). The questions and images are in the same order as in the survey. There are 15 questions concerned with the reconstruction accuracy. For each image, we provide the metric used to determine the best (or worst) match alongside the method used to obtain that result (obviously, such information was not revealed to the respondents).

1. Which of the following images presents the urban area in the MOST DETAILED WAY?

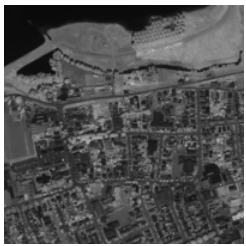
Reference image

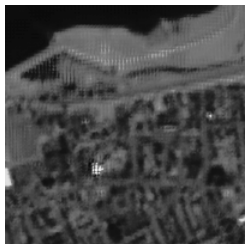
PSNR (HRN_SIM), 21 votes

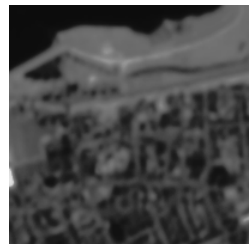
SSIM (Lanczos), 9 votes

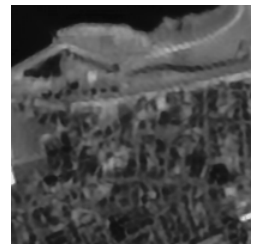
LPIPS (RAMS_NIR), 136 votes

2. Which of the following images is the MOST DIFFERENT from the reference image?

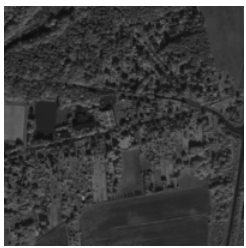
Reference image

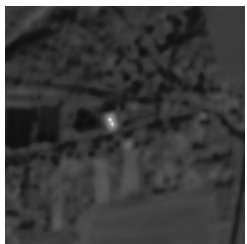
LPIPS (Bicubic), 119 votes

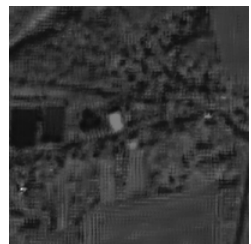
SSIM (HRN_SIM), 29 votes

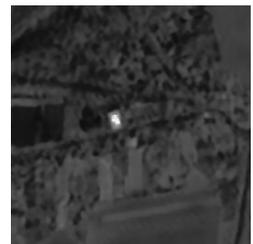
PSNR (HRN_NIR), 18 votes

1

Table 1 – *Continued from the previous page*

3. Which of the following images presents the annotated woodless area in the MOST ACCURATE way?
   This area was annotated in red in the reference image (right).

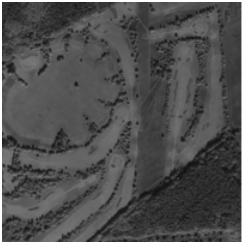 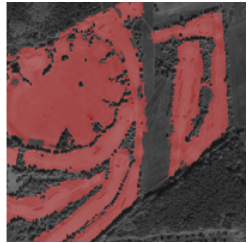 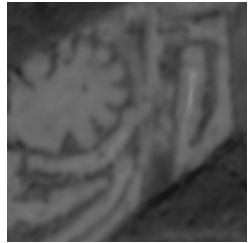 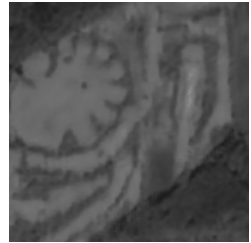 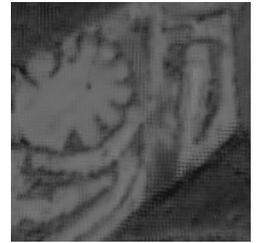

| Reference image | SSIM (Lanczos), 10 votes | LPIPS (HRN_NIR), 89 votes | PSNR (RAMS_SIM), 67 votes |

4. Which of the following images presents the SHAPE of annotated buildings in the WORST way?
   The buildings of interest are annotated in red in the reference image.

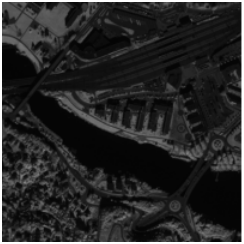 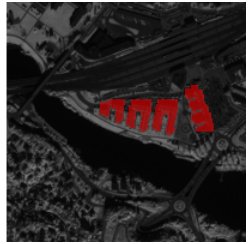 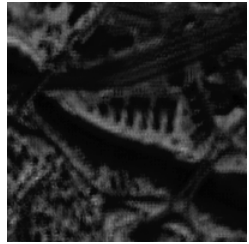 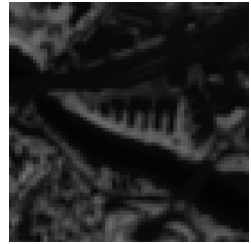 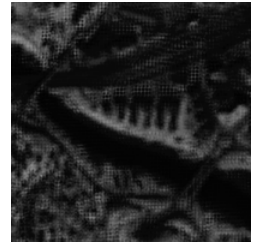

| Reference image | PSNR (HRN_SIM), 10 votes | LPIPS (NN), 139 votes | SSIM (RAMS_SIM), 17 votes |

5. In which image the rural roads are presented in the CLEAREST (MOST DETAILED) way?
   The roads of interest are annotated in red in the reference image.

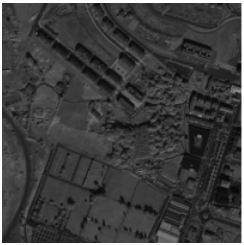 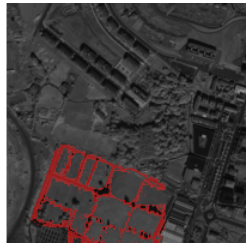 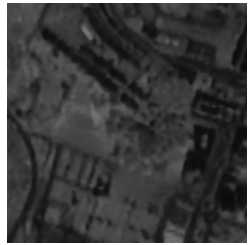 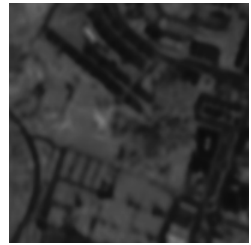 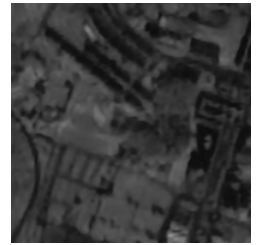

| Reference image | LPIPS (HRN_NIR), 69 votes | PSNR (Lanczos), 21 votes | SSIM (RAMS_NIR), 76 votes |

Table 1 – *Continued from the previous page*

6. In which image the buildings manifest THE LEAST LEVEL OF DETAIL?

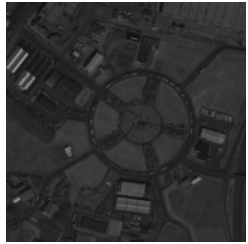
Reference image

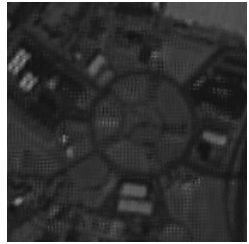
LPIPS (RAMS_SIM), 78 votes

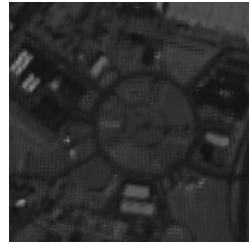
SSIM (HRN_SIM), 50 votes

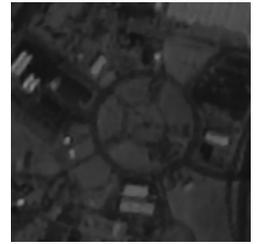
PSNR (HRN_NIR), 38 votes

7. In which of the following images the junction is presented in the MOST ACCURATE (DETAILED) way?

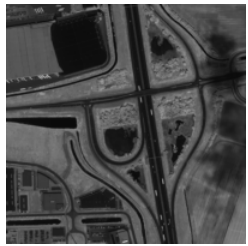
Reference image

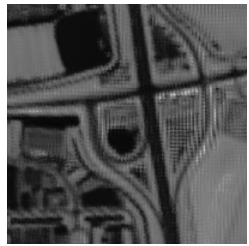
PSNR (RAMS_SIM), 30 votes

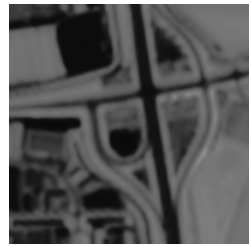
SSIM (Lanczos), 34 votes

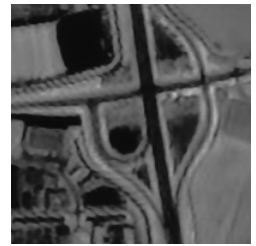
LPIPS (RAMS_NIR), 102 votes

8. In which of the following images it is the MOST CHALLENGING to distinguish separate trees?

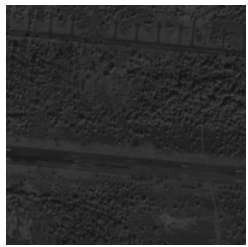
Reference image

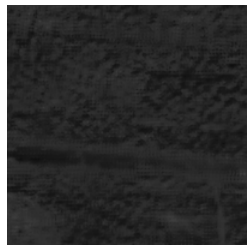
SSIM (RAMS_SIM), 12 votes

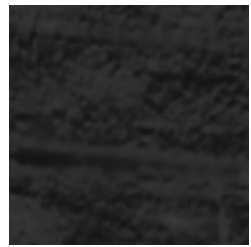
LPIPS (Bicubic), 150 votes

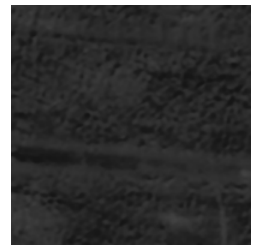
PSNR (RAMS_NIR), 4 votes

9. In which of the following images the area of interest (annotated in red in the Reference image) is presented in the MOST DETAILED way?

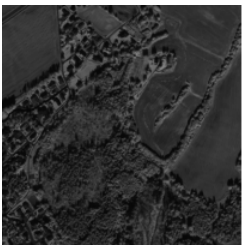
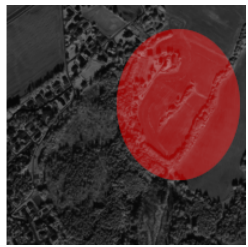
Reference image

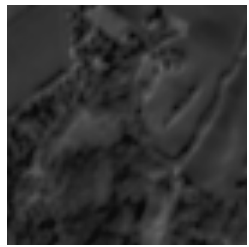
PSNR (Linear), 3 votes

LPIPS (RAMS_NIR), 150 votes

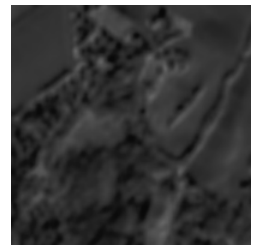
SSIM (Bicubic), 13 votes

Table 1 – *Continued from the previous page*

10. Which of the following images presents the coastal area in the LEAST ACCURATE (LEAST DETAILED) way?
    The area of interest is rendered in red in the reference image.

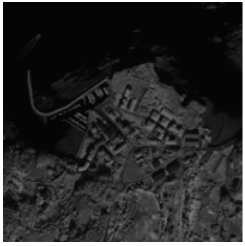 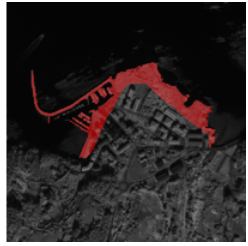 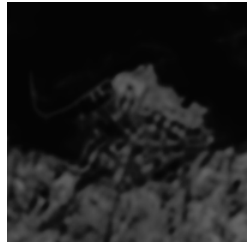 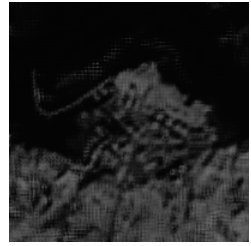 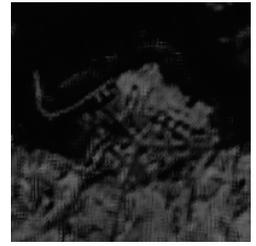

| Reference image | | LPIPS (Lanczos), 154 votes | SSIM (RAMS_SIM), 2 votes | PSNR (HRN_SIM), 10 votes |

11. Which of the following images presents the roads in the MOST DETAILED way?
    The roads of interest are annotated in red in the reference image.

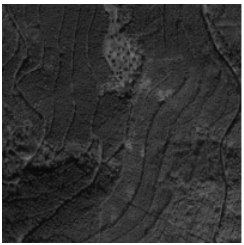 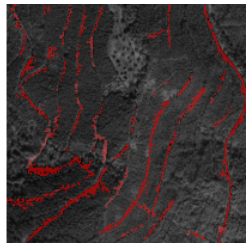 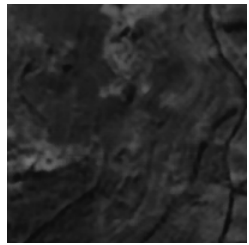 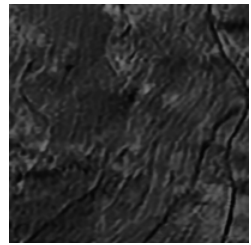 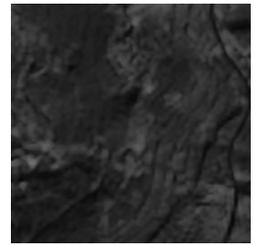

Reference image — SSIM (HRN_RED), 10 votes — LPIPS (RAMS_NIR), 126 votes — PSNR (RAMS_RED), 30 votes

12. In which of the following images the urban area is presented in the LEAST DETAILED way?

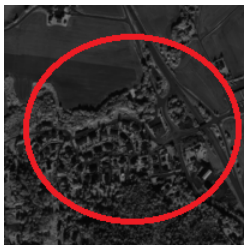 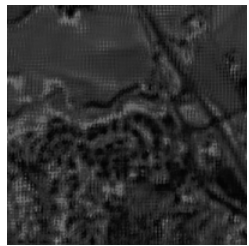 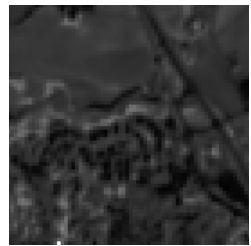 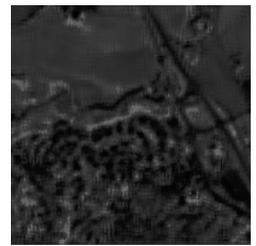

Reference image — SSIM (RAMS_SIM), 11 votes — LPIPS (NN), 151 votes — PSNR (HRN_SIM), 4 votes

Table 1 – *Continued from the previous page*

13. Which of the following images allows one to count the buildings in THE MOST CONVENIENT way?
    The buildings of interest are annotated in red in the reference image.

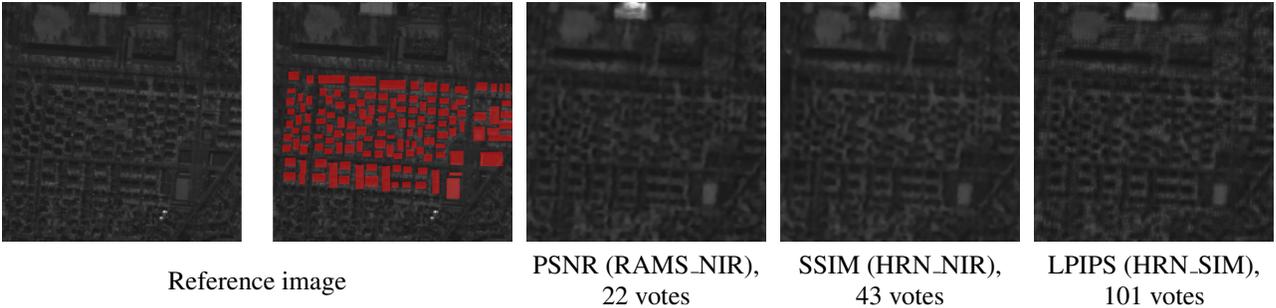

Reference image  |  PSNR (RAMS_NIR), 22 votes  |  SSIM (HRN_NIR), 43 votes  |  LPIPS (HRN_SIM), 101 votes

14. In which of the following images the parking is presented in the LEAST DETAILED way?
    The parking of interest is annotated in red in the reference image.

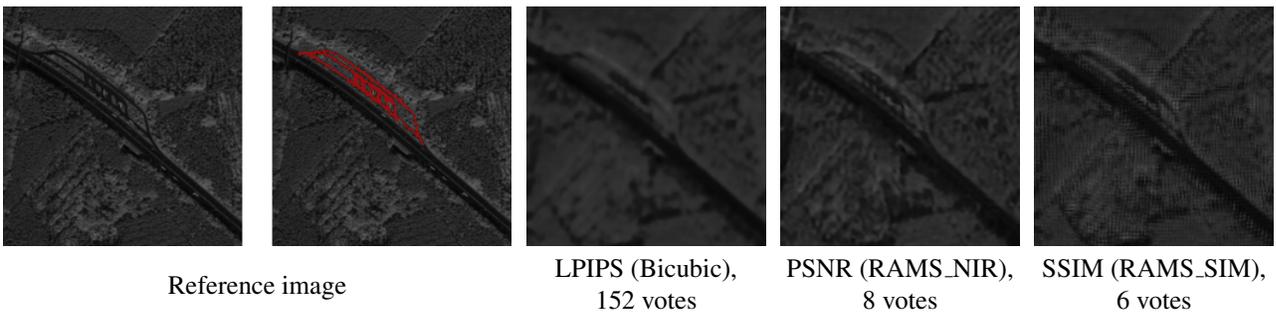

Reference image  |  LPIPS (Bicubic), 152 votes  |  PSNR (RAMS_NIR), 8 votes  |  SSIM (RAMS_SIM), 6 votes

15. Which of the following images presents the roads of interest MOST FAITHFULLY?
    The roads of interest are annotated in red in the reference image.

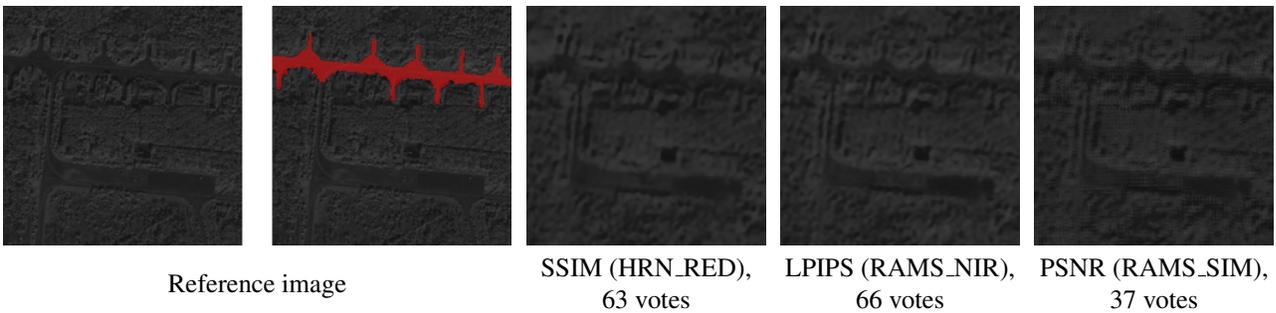

Reference image  |  SSIM (HRN_RED), 63 votes  |  LPIPS (RAMS_NIR), 66 votes  |  PSNR (RAMS_SIM), 37 votes



\end{document}